\documentclass[acmsmall,screen]{acmart}

\AtBeginDocument{%
  \providecommand\BibTeX{{%
    \normalfont B\kern-0.5em{\scshape i\kern-0.25em b}\kern-0.8em\TeX}}}

\usepackage{multirow,makecell}
\usepackage{tcolorbox}
\usepackage{color,xcolor}
\usepackage{listings,amsfonts}
\usepackage{caption}
\usepackage{subcaption}
\usepackage{threeparttable}
\usepackage{bbding}
\usepackage{graphicx}
\usepackage{booktabs} 
\usepackage{longtable}
\usepackage{array}
\usepackage{url}
\usepackage{bbding}
\usepackage{pifont}

\definecolor{customblue}{HTML}{006ca6}

\AtEndPreamble{
	\usepackage{hyperref}
	\hypersetup{
		colorlinks = true,
		linkcolor = customblue,
		anchorcolor = customblue,
		citecolor = customblue,
		filecolor = customblue,
		urlcolor = customblue
	}
}

\usepackage{tcolorbox}
\def\BibTeX{{\rm B\kern-.05em{\sc i\kern-.025em b}\kern-.08em
    T\kern-.1667em\lower.7ex\hbox{E}\kern-.125emX}}

\begin{document}

\setcopyright{acmlicensed}
\acmJournal{TOSEM}
\acmYear{2026} \acmVolume{35} \acmNumber{9} \acmArticle{262} \acmMonth{9}\acmDOI{10.1145/3819579}

\title{Mobile App Analysis in the New Era: Challenges and the Road Ahead}

\author{Jiale Wu}
\email{jialewu@hust.edu.cn}
\orcid{0009-0007-5682-2950}
\authornotemark[1]
\affiliation{%
  \department{Hubei Key Laboratory of Distributed System Security}
  \department{Hubei Engineering Research Center on Big Data Security}
  \department{School of Cyber Science and Engineering}
  \institution{Huazhong University of Science and Technology}
  \city{Wuhan}           
  \country{China}
}

\author{Tianming Liu}
\email{tmliu@hust.edu.cn}
\orcid{0000-0002-5216-933X}
\authornote{Jiale Wu and Tianming Liu are the co-first authors.}
\affiliation{%
  \department{Hubei Key Laboratory of Distributed System Security}
  \department{Hubei Engineering Research Center on Big Data Security}
  \department{School of Cyber Science and Engineering}
  \institution{Huazhong University of Science and Technology}
  \city{Wuhan}           
  \country{China}
}

\author{Yanjie Zhao}
\email{yanjie_zhao@hust.edu.cn}
\orcid{0000-0001-8793-5367}
\affiliation{%
  \department{Hubei Key Laboratory of Distributed System Security}
  \department{Hubei Engineering Research Center on Big Data Security}
  \department{School of Cyber Science and Engineering}
  \institution{Huazhong University of Science and Technology}
  \city{Wuhan}           
  \country{China}
}

\author{Haoyu Wang}
\email{haoyuwang@hust.edu.cn}
\authornote{Haoyu Wang is the corresponding author (haoyuwang@hust.edu.cn).}
\orcid{0000-0003-1100-8633}
\affiliation{
  \department{Hubei Key Laboratory of Distributed System Security}
  \department{Hubei Engineering Research Center on Big Data Security}
  \department{School of Cyber Science and Engineering}
  \institution{Huazhong University of Science and Technology}
  \city{Wuhan}           
  \country{China}
}

\begin{abstract}
Mobile applications (apps) have grown into complex, multi-platform systems that support diverse services and operate within varying regulatory and distribution contexts. Traditional analysis techniques have established a strong foundation for ensuring quality, security, and compliance, but face increasing challenges as mobile apps evolve. This paper presents a three-layer perspective to examine these challenges. The Artifact Layer highlights how heterogeneous code bases and protection mechanisms reduce the analyzability of packaged apps. The Runtime Layer addresses growing execution complexity, from multi-modal interactions and super-app architectures to AI-driven autonomous operations. The Ecosystem Layer considers broader contexts including emerging platforms, fragmented distribution channels, and region-specific regulatory requirements. New characteristics across these layers interact and compound, creating difficulties that existing approaches are not designed to handle. For each layer, we identify key challenges and potential research directions, including framework-aware analysis, adaptive testing, and policy-aware compliance verification. We also discuss how large language models bring new analytical capabilities to the field while their integration into apps raises new security and privacy concerns. This paper aims to clarify where established methods fall short and highlight directions for future research in this evolving field.
\end{abstract}



\maketitle

\clearpage

\section{Introduction}

The rapid evolution of mobile technology has fundamentally transformed how people interact with digital devices, making mobile apps common in daily life. These apps now range from simple utilities to complex, platform-integrated systems that handle sensitive data, financial transactions, and communications. This requires methods that can ensure quality, security, and performance at scale, establishing mobile app analysis as an important field of research and development.
In this paper, we use the term mobile app analysis to refer to the systematic examination of mobile apps to evaluate their implementation and behavior through automated and semi-automated techniques. Established approaches include static analysis, which inspects code implementations within app packages to recover structure and potential problems (e.g., Flowdroid~\cite{arzt2014flowdroid}); dynamic analysis, which observes runtime behavior to assess actual functionality and resource usage (e.g., Sapienz~\cite{mao2016sapienz}); and hybrid approaches, which combine static and dynamic techniques for more comprehensive evaluation. These techniques have established a strong foundation and are widely used in both academic research and industrial practice (as introduced in \S\ref{sec:mobile-app-analysis}).

However, contemporary mobile apps are increasingly reshaped by rapid technological advances, evolving development paradigms, and changing platform constraints. Modern development practices have introduced new architectural patterns (e.g., cross-platform frameworks such as React Native~\cite{nodejs_rn,Chen24MOBILESoft} and Flutter~\cite{flutter}) and execution paradigms (e.g., super-app ecosystems~\cite{DBLP:journals/tem/SalehiMNG24,DBLP:conf/uss/WangZL23} and agent-driven interactions~\cite{appagent,mobileagentv2}) that create novel forms of complexity. These changes alter not only what needs to be analyzed but also how analysis can be effectively performed, leaving traditional approaches with blind spots. As a result, gaps have emerged between the capabilities of current analysis methods and the realities of modern apps~\cite{arzt2014flowdroid,mao2016sapienz,pan2020QTesting}.

To address these challenges systematically, we introduce a layered perspective that organizes emerging features of mobile apps and their analysis challenges by scope: the \textbf{Artifact Layer} (the packaged app and its code and resources), the \textbf{Runtime Layer} (the app's behavior and interactions during execution), and the \textbf{Ecosystem Layer} (the surrounding platform, distribution, and governance environment). By moving from app artifacts outward to ecosystem constraints, this framework clarifies how different types of complexity interact and compound across layers.

At each layer, we examine emerging changes, the challenges they create, and potential research directions. 
Taken together, our analysis indicates that the evolution of mobile apps is driven not by isolated changes, but by changes across artifacts, runtime, and ecosystems that interact and reinforce one another. 
Heterogeneous code bundles, more complex execution behaviors, and fragmented deployment contexts combine to produce forms of complexity that existing techniques struggle to address. 
The layered framework highlights key challenges: reduced analyzability in packaged artifacts, increased runtime complexity from super-app architectures, multi-modal interactions, and agent-driven operations, and greater difficulty in maintaining consistency across diverse platforms and regulatory contexts. 
Based on these observations, we outline research directions that combine platform-specific adaptation with broader unifying methods, aiming to sustain the effectiveness of mobile app analysis as the ecosystem continues to evolve.

The remainder of this paper is structured as follows. 
Section~\ref{sec:layered-perspective} reviews current mobile app analysis methods and introduces the three-layer perspective. 
Section~\ref{sec:artifact-layer}--\ref{sec:ecosystem-layer} then analyze the Artifact, Runtime, and Ecosystem Layers in detail, each structured around emerging characteristics, analysis challenges, and possible research directions. 
Section~\ref{sec:discussion} brings together the findings across layers, compares academic and industrial viewpoints, and highlights trends and insights in the evolution of mobile app analysis. Finally, Section~\ref{sec:conclusion} concludes the paper.

\section{Mobile App Analysis: Current Methods and a Layered Perspective}
\label{sec:layered-perspective}

Mobile app analysis has made significant progress in specific areas, but as apps evolve, they become more powerful yet increasingly difficult to analyze thoroughly and effectively. This section first reviews current mobile app analysis techniques (\S\ref{sec:mobile-app-analysis}), then introduces our three-layer perspective to observe the landscape of mobile apps and organize the challenges (\S\ref{sec:3-layer-perspective}).

\subsection{Overview of Mobile App Analysis}
\label{sec:mobile-app-analysis}

This section provides an overview of static and dynamic methods, hybrid approaches, and AI-integrated solutions that constitute the main technical foundations of mobile app analysis. These techniques have been developed over many years and are now widely adopted in both academic research and industrial practice. They form the established toolkit that researchers and practitioners rely on for examining mobile apps.

\textbf{Static Analysis Techniques.} Static analysis operates by examining app implementations in bytecode or binary form without execution, typically using frameworks such as Soot~\cite{lam2011soot} and WALA~\cite{wala2024tj}. It provides comprehensive coverage by reasoning about all possible control and data-flow paths, and has supported a wide range of security and reliability tasks. Representative usages include the detection of private data leaks~\cite{arzt2014flowdroid,li2015iccta,Rasthofer201440,li2014automatically,wei2018amandroid,gibler2012androidleaks,lu2012chex,chan2012droidchecker,zhang2014efficient}, identification of vulnerabilities~\cite{luoshi2013a3,vidas2014a5,wei2018amandroid,androguard2012,androlizer2011,DBLP:conf/icse/ChenTPL25}, enforcement of permission checks~\cite{cui2015wechecker,yang2015appcontext,Bartel2014617permission}, detection of malware~\cite{PanGFF20SLRMalSta,Ma19Combination,Zhang19Efficient,Ge19AMDroid,Sercan19AndMFC,Shen19Android,Xu19AndroidCFG,Fan19Graph,DBLP:journals/pacmse/LiCZXXW25,DBLP:journals/pacmse/LiCZXXW25a}, and identification of app clones~\cite{chen2014clone,lindorfer2014andradar}. A persistent challenge is the increasing heterogeneity of code bases: modern apps often combine managed code (Dalvik/ART bytecode) with native binaries, requiring cross-language analysis techniques~\cite{samhi2022jucify,liu2023reunify,wang2024nativesummary} to bridge semantic gaps. While static analysis offers breadth and completeness, it often generates false positives and incurs high computational cost, especially on large or heavily obfuscated apps.

\textbf{Dynamic Analysis Techniques.} Dynamic analysis seeks to establish actual behavior by executing apps in controlled environments and observing their interactions with users, system services, and external resources. Representative frameworks such as Monkey~\cite{monkeyGoogle}, Sapienz~\cite{mao2016sapienz}, Stoat~\cite{su2017Stoat}, Droidbot~\cite{DroidBot}, and QTesting~\cite{pan2020QTesting} illustrate this trajectory, with LLM-driven approaches~\cite{liu2024make,DBLP:journals/pacmse/WangLZYW25} representing recent advances. Tools for simulating user interactions~\cite{pilgun2020fineAVCTOOL,dong2020timemachine,wang2021infrastructureWCTester,romdhana2022deepAres,almeida2018chimp,choudhary2015AndroTest,li2019humanoid,DBLP:journals/pacmse/LiuTLCRPD25} have further improved coverage~\cite{DBLP:conf/icse/AkinotchoWR25}, GUI exploration~\cite{DBLP:conf/icse/Zhang0XL025,DBLP:conf/icse/MehralianHM25}, and fault detection~\cite{DBLP:conf/icse/GuoWD00ZCLY25}. Beyond testing, dynamic analysis underpins security and privacy assessments: runtime evidence such as network traffic~\cite{wu2019openports,wang2020identifyingSSL,liu2018automaticallySSL,shi2019mossot,baskaran2023authleak,li2023usertagspoofing}, system calls~\cite{Mario2019DyMal,DLDroid2020,DAngeloFP20Malware}, and memory snapshots~\cite{ZhangWHL20DAMBA,BenzKLBBZ21,Yang22FSAFlow} enables compliance evaluation~\cite{ReardonFWOVE19,PapageorgiouSPA18Health,Nguyen0MS21GDPR,AndowALESX17UiRef,LiuLZWZ20Privacy,DiamantarisPMIP19REAPER} and malicious behavior detection~\cite{Mario2019DyMal,DLDroid2020,CaiMRY19DroidCat,ArshadSWMSY18SAMADroid,XuLDC18DeepRefiner,FengMSXM18DynamicMal,AhmadCCBZ20StaDART,GajraniALBGTZ20EspyDroid,LorenzoMMMS20VizMal,ZhangWHL20DAMBA,CaiFH20runtime,BhatBD23malware,Heng23Black-box}. It is particularly valuable for capturing runtime-only phenomena such as dynamically loaded code~\cite{AhmadCCBZ20StaDART} or unauthorized data exfiltration~\cite{tam2015copperdroid}, though limited by incomplete coverage, nondeterministic behavior, and the difficulty of reproducing complex execution environments.

\textbf{Hybrid Analysis Techniques.} Recognizing the complementary strengths and weaknesses of static and dynamic methods, hybrid approaches integrate the two. A common pattern is to use static analysis to flag potential data-flow paths and then validate them dynamically~\cite{Khaled23ViaLin,Yang22FSAFlow}, or conversely to use dynamic traces to refine or confirm static signatures. Hybrid techniques have been especially influential in malware detection~\cite{DLDroid2020,FengMSXM18DynamicMal,CaiMRY19DroidCat} and taint analysis~\cite{AhmadCCBZ20StaDART,GajraniALBGTZ20EspyDroid,CaiFH20runtime}, where neither static nor dynamic alone is sufficient. The main challenge lies in designing effective integration strategies that balance coverage, precision, and scalability.

\textbf{AI-integrated Analysis Solutions.} Recent years have seen the rapid infusion of artificial intelligence into mobile app analysis, leveraging traditional machine learning (ML), deep learning (DL), and increasingly, large language models (LLMs). 
ML-based methods have long been established in malware classification~\cite{Arp2014Drebin,Mariconti2017MaMaDroid} and code similarity detection. For instance, statistical features are used to identify third-party libraries~\cite{Ma2016LibRadar} and analyze malware lineage~\cite{Fan2021MalRadar}. 
To enhance automation, DL and reinforcement learning (RL) have been introduced to guide input generation~\cite{pan2020QTesting} and learn user interaction patterns~\cite{li2019humanoid} for testing. 
More recently, LLMs have extended these capabilities by leveraging semantic understanding: they have been applied to reproduce crashes from bug reports~\cite{Icse2025crashtranslator}, translate natural language into GUI operations~\cite{Feng2024AdbGPT}, and assist in reverse engineering tasks such as deobfuscation~\cite{Bichsel2016DeGuard}. 
While surveys on LLM for software engineering~\cite{Fan2023LLMSurvey} highlight their potential, effectively combining the reasoning power of LLMs with the precision of static and dynamic analysis remains a critical challenge to ensure robustness and explainability. In subsequent sections, we will explore the impact of LLM on mobile app analysis, with a focused discussion on this topic in \S\ref{sec:discussion}.

\subsection{The Three-Layer Perspective}
\label{sec:3-layer-perspective}

\begin{figure}
  \centering
  \includegraphics[width=1.0\linewidth]{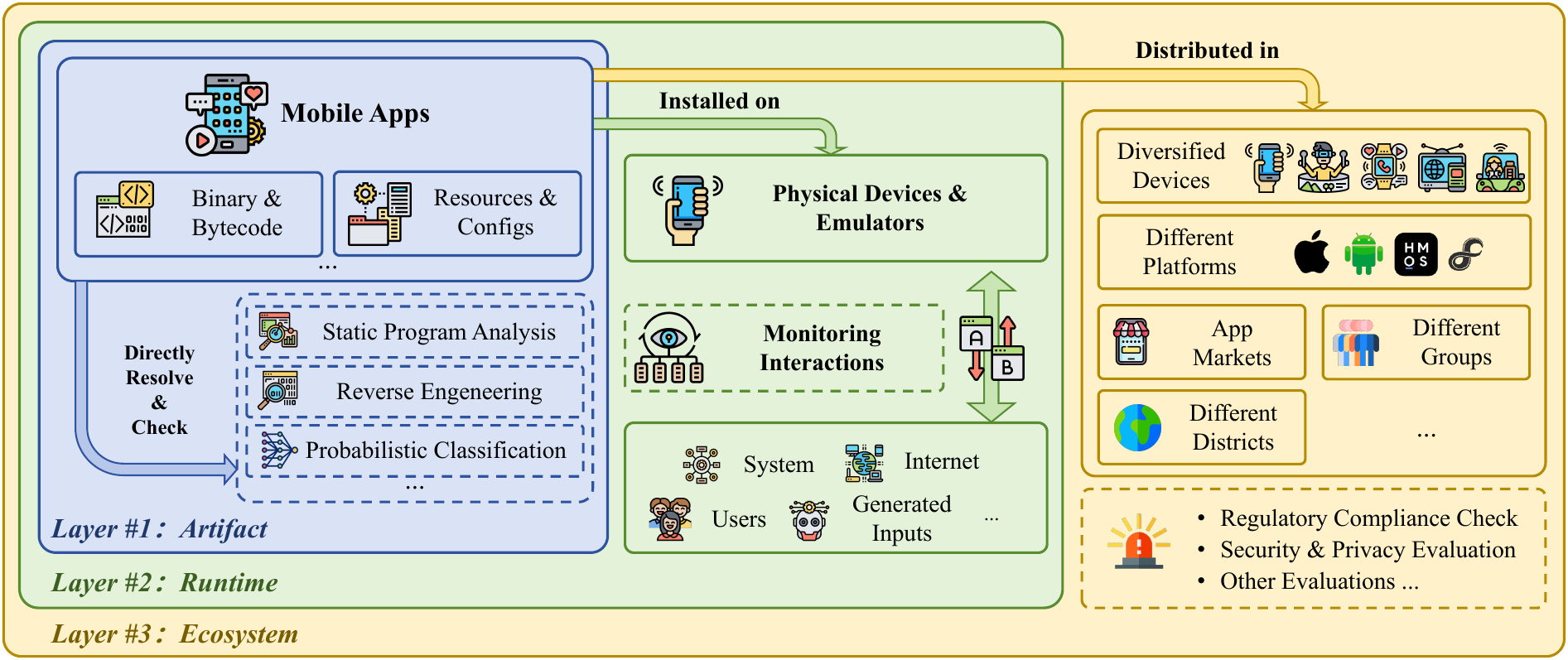}
  \caption{Three-Layer Perspective for Examining Mobile App Analysis Challenges}
  \label{fig:3-layer-overview}
\end{figure}

Mobile app analysis covers diverse techniques as introduced above, while the mobile ecosystem has evolved across multiple dimensions. To systematically examine how these changes affect analysis capabilities, we present a three-layer perspective that organizes them by scope—starting from the app package itself and progressively moving outward to its surrounding environment. \autoref{fig:3-layer-overview} illustrates this organization. At the Artifact Layer, we focus on packaged apps as distributed bundles—the code, resources, and structures that analysis tools need to first process. At the Runtime Layer, we examine apps during execution, observing how they behave and interact in operational contexts. At the Ecosystem Layer, we consider the broader environment where apps are distributed and regulated, including diverse platforms, markets, and deployment frameworks.

This organization allows us to systematically relate changes in mobile apps to their effects on analysis capabilities. By examining each layer in turn, we can identify where traditional approaches encounter difficulties and where new capabilities are needed.

\begin{itemize}
  \item \textbf{Artifact Layer — \textit{Changes in packaged apps}}: 
  This innermost layer considers the app as a packaged artifact, focusing on how its internal composition influences what can be recovered and examined. Mobile app evolution has introduced heterogeneous components packaged in a bundle and increasingly sophisticated protection mechanisms, which obscure or transform implementation logic and thereby reduce transparency. These shifts challenge the ability of analysis techniques (static or otherwise) to reason about code structure, detect data leaks~\cite{arzt2014flowdroid,wei2018amandroid}, and assess security properties.

    \item \textbf{Runtime Layer — \textit{Changes during execution}}: 
    Moving outward from the packaged artifact, this layer focuses on the app in action: how it behaves when executed in real environments and how it interacts with users, system services, and external resources. Modern apps exhibit increasingly complex runtime behaviors that diverge from their static representations, including adaptive functionality, dynamic content loading, and sophisticated interaction patterns. These characteristics affect both monitoring and testing, requiring new strategies that go beyond traditional frameworks~\cite{mao2016sapienz,pan2020QTesting,liu2024make}.

  \item \textbf{Ecosystem Layer — \textit{Changes in platform context}}: 
  Beyond the app itself and its runtime, analysis is expected to also account for the broader ecosystem in which apps are developed, distributed, and regulated. This outermost layer defines the boundary for what analysis should target and verify, covering platform diversification, distribution fragmentation, and regulatory divergence. Such ecosystem-level factors shape compliance objectives and evidence standards, constraining and guiding analysis at the Artifact and Runtime Layers. By situating these external forces as the outer scope, this layer completes the inside-out narrative and highlights opportunities that arise when technical analysis is integrated with contextual awareness. 
\end{itemize}

This layered perspective enables us to systematically analyze mobile app challenges by moving from technical implementation details to broader ecosystem considerations.

\section{Artifact Layer}
\label{sec:artifact-layer}

Our analysis starts with the app as a packaged artifact, focusing on the code in the distributed bundle and what can be inferred from it. Packaging and build practices shape the app’s observable surface, deciding what remains visible from development and how logic is split across representations. We then outline the key characteristics of packaged artifacts, the challenges they pose, and possible ways to address them.

\begin{figure}[h]
  \centering
  \includegraphics[width=1.0\linewidth]{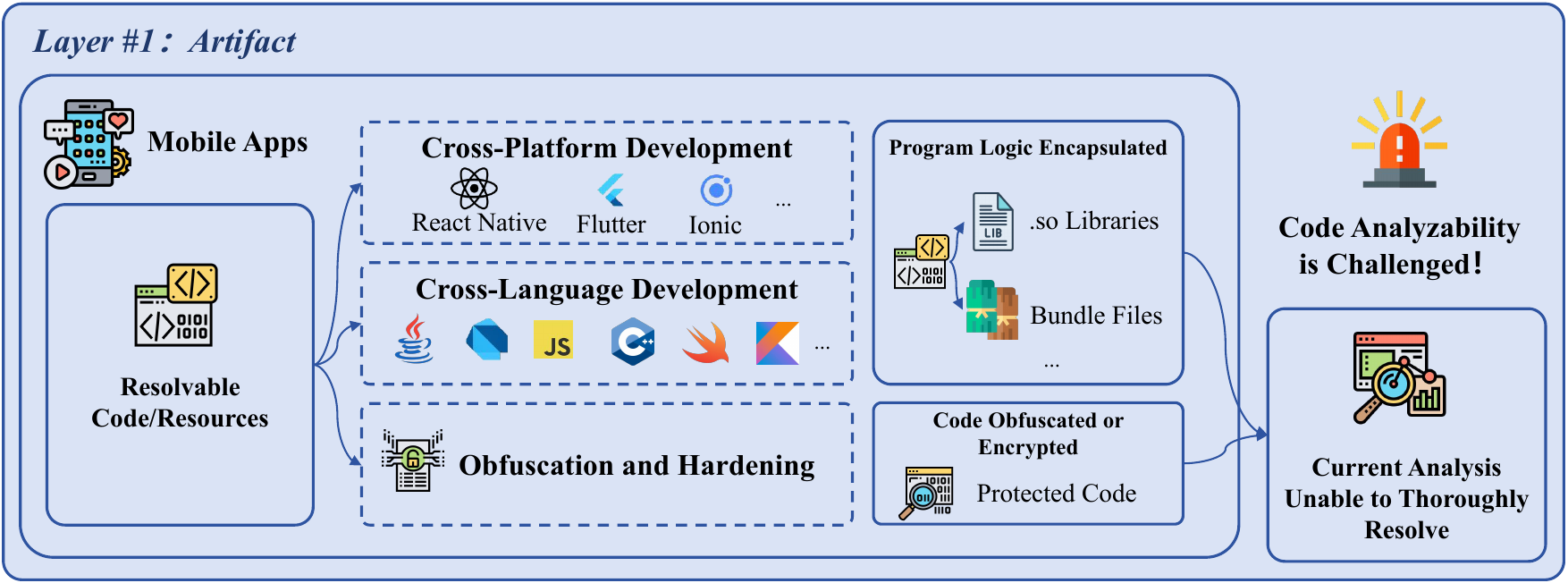}
  \caption{Overview of Artifact Layer}
  \label{fig:artifact-layer-fig}
\end{figure}

\subsection{Emerging Characteristics}
At the Artifact Layer, the representation of implementation logic in packaged apps has shifted from uniform, analyzable bytecode to heterogeneous, framework-shaped, and protection-hardened artifacts (as illustrated in \autoref{fig:artifact-layer-fig}). This section highlights the main trends that redefine what can be observed from packaged apps.

\textbf{Mixed Code Architectures.} Contemporary mobile apps rarely remain within a single language or runtime. Instead, they integrate managed code (e.g., Java/Kotlin bytecode) with precompiled native libraries and other components that contain implementation logic. Two specific directions dominate this trend:

\begin{itemize}

\item \textbf{Cross-Platform Development.} The adoption of cross-platform frameworks (e.g., React Native~\cite{nodejs_rn}, Flutter~\cite{flutter}, Ionic~\cite{ionic}, and Kotlin Multiplatform~\cite{kotlin_multiplatform}) has changed mobile development by enabling single-codebase deployment across platforms. A recent empirical analysis highlights their prevalence~\cite{Chen24MOBILESoft}: React Native appears in 14 of the top 100 apps on Google Play, while Flutter has 39 implementations in China's Tencent App Store. These frameworks differ significantly in how they package app logic: Kotlin Multiplatform leverages Java interoperability~\cite{kotlin} to produce analyzable Dalvik bytecode, while Flutter compiles Dart code~\cite{dart} into native binaries (libapp.so), and React Native bundles JavaScript code executed through bridge mechanisms~\cite{nodejs_rn}.

\item \textbf{Multi-Language Integration.} Even within a single platform, apps increasingly combine multiple languages. Android apps mix Java/Kotlin with C/C++ through the NDK, iOS apps integrate Swift with Objective-C, and many apps embed JavaScript or Lua for dynamic features. Domain-specific languages further specify resources and UI behavior. App logic thus spans Dalvik/ART bytecode, native modules, and embedded scripts within single packages.\looseness=-1

\end{itemize}

\textbf{Obfuscation and Protection.} Packaging and protection mechanisms have advanced to deliberately reduce artifact transparency. Despite being a recognized challenge for over a decade, obfuscation adoption continues to grow. Common techniques include identifier renaming, string encryption and control-flow transformations~\cite{dong2018obfuscation,DBLP:conf/icse/WangBWWCWW18}. More advanced approaches use virtualization-based obfuscation, compiling code into custom bytecode executed by generated interpreters~\cite{zhao2020virtualization,sok_vmobf}. Packing services encrypt binaries and insert loader stubs with anti-analysis logic~\cite{duan2018packers}, while runtime defenses such as anti-debugging, emulator detection, and integrity checks are widely deployed~\cite{berlato2020adoption,arap2024}. Commercial protection services including 360 Jiagu~\cite{jiagu360_2025}, Tencent Legu~\cite{tencent_ms_2025}, Alibaba MSA~\cite{aliyun_msa_2025}, Appdome~\cite{appdome_2025}, and Promon~\cite{promon_2025} offer comprehensive hardening solutions. Empirical studies show obfuscation usage in Google Play increased by 13\% over the last decade~\cite{obf_practices_2025}, and sophistication continues to evolve~\cite{Ebad24AndroidObDeOb,DBLP:journals/csur/SutterSCK25} despite ongoing deobfuscation research~\cite{DBLP:journals/eswa/PatsakisCL24,DBLP:conf/dimva/BesteMHFCBHES25}.

Together, cross-platform development, multi-language composition, and protection mechanisms reduce the analyzability of packaged apps and increase analysis difficulty.

\subsection{Analysis Challenges}

The changes described above create fundamental challenges for mobile app analysis at the artifact level. The core problem is loss of analyzability: implementation logic that was previously accessible to direct inspection has become difficult to parse and understand. Traditional static analysis techniques, designed for parsing bytecode into intermediate representations and performing analysis on them~\cite{arzt2014flowdroid,lam2011soot,wala2024tj}, now face complexity that manifests in two interconnected ways.

\textbf{Semantic Opacity and Abstraction Barriers.} Cross-platform frameworks and obfuscation techniques combine to prevent clear analysis of packaged apps. Frameworks introduce abstraction layers that embed logic in non-standard representations: React Native uses JavaScript bundles executed via bridges~\cite{nodejs_rn}, Flutter compiles Dart to native binaries~\cite{flutter}, each requiring different analysis approaches than traditional bytecode inspection. These framework-specific transformations create barriers that conventional decompilation struggles to penetrate. Tools designed for bytecode have difficulty interpreting these components. Obfuscation compounds these barriers by deliberately obscuring whatever structure remains accessible. Virtualization-based techniques~\cite{zhao2020virtualization,sok_vmobf} wrap code in custom interpreters, control-flow transformations~\cite{dong2018obfuscation} obscure program logic, and multi-layer protection schemes combine multiple mechanisms to create artifacts that appear syntactically valid but remain semantically opaque. The result is that automated tools often fail to recover the program semantics needed for compliance verification. The additional abstraction layers create fundamental gaps in what can be analyzed.

\textbf{Heterogeneous Composition.} Modern apps present fragmented analysis surfaces spanning multiple languages, runtimes, and protection schemes within single packages. Existing analysis techniques have been designed to handle specific representations effectively. Static analysis frameworks are designed for examining bytecode~\cite{wala2024tj,lam2011soot}, native code analyzers can dissect native binaries~\cite{samhi2022jucify,wang2024nativesummary}. However, no single technique provides comprehensive coverage across heterogeneous compositions. In the foreseeable future, methods for analyzing various components of app artifacts are expected to emerge. It is a challenge to integrate these methods to form a comprehensive methodology.

\subsection{Potential Research Directions}

The challenges outlined above point toward research directions that could improve analytical capabilities for modern mobile apps. These directions address two fundamental problems: recovering semantics from obscured artifacts and handling heterogeneous code compositions.

\textbf{Recovering Semantics from Framework Abstractions and Obfuscation.} Cross-platform frameworks and protection mechanisms both reduce analyzability, but through different means. Frameworks introduce abstraction layers that embed code in non-standard representations: JavaScript bundles in React Native, Dart-compiled binaries in Flutter, and bridge mechanisms connecting framework code to native implementations.
Analyzing these requires framework-aware techniques that understand how each framework compiles, packages, and executes code~\cite{nodejs_rn,flutter,dart}. Existing analysis work for native binaries or React Native bundles~\cite{samhi2022jucify,liu2023reunify,wang2024nativesummary} provides foundations, but framework-specific abstractions require specialized handling beyond general multi-language support. Protection mechanisms deliberately obscure code through obfuscation and encryption. Addressing this requires approaches that can adapt to evolving strategies rather than targeting specific techniques. AI-assisted methods show promise here: neural networks have been applied to defeat virtualization-based obfuscation~\cite{DBLP:journals/tifs/ZhaoZKPDZX25}, while LLM-based approaches demonstrate capabilities in code deobfuscation~\cite{DBLP:journals/corr/abs-2506-20170,DBLP:conf/dimva/BesteMHFCBHES25}, reverse engineering~\cite{hu2024degpt,tan2024llm4decompile}, symbol recovery~\cite{xie2024resym}, and control flow reconstruction~\cite{DBLP:journals/eswa/PatsakisCL24}. These techniques could create analysis tools that evolve alongside protection mechanisms~\cite{DBLP:conf/icyberphys/GolovkoSVKS24,llm_obfuscate_2024}, though their robustness and reliability require further validation.

\textbf{Unified Analysis Across Heterogeneous Compositions.}  
Modern apps combine Java/Kotlin bytecode, C/C++ native libraries, JavaScript or Dart modules, and domain-specific resource languages within single bundles, fragmenting the analysis surface. Existing efforts such as lifting native code into Jimple IR~\cite{samhi2022jucify,liu2023reunify} or summarizing native functions for bytecode analyzers~\cite{wang2024nativesummary} show that cross-language reasoning is possible, but current solutions remain limited to narrow integration scenarios. Industrial practice suggests that broader unification is achievable. Ant Group’s YASA Security Station~\cite{AntGroupYASA2025}, for example, integrates analyses across binaries, network logs, and host/cloud modules, orchestrating heterogeneous tools while aligning outputs into unified insights. Though designed for enterprise security rather than mobile apps, it demonstrates the feasibility of modular integration at scale. Inspired by this, we argue that mobile app analysis should pursue unified approaches that go beyond pairwise adapters. Promising directions include translating heterogeneous representations into a shared intermediate form, or developing new multi-language IRs that retain component-specific semantics while supporting whole-app reasoning. Equally important are modular architectures that can plug in specialized analyzers while tracking dependencies and control/data flows consistently across runtimes. Such approaches would provide a principled way to analyze the increasingly diverse compositions found in modern app bundles.

\section{Runtime Layer}
\label{sec:runtime-layer}

The Runtime Layer shifts the focus from packaged artifacts to the behavior of apps during execution. At this layer, we examine how apps operate in runtime environments, interact with users and system services, and adapt their functionality. Modern mobile apps exhibit increasingly complex runtime characteristics that diverge from their static representations, challenging previous analysis approaches that were designed for more predictable execution patterns.

\subsection{Emerging Characteristics}

\begin{figure}
  \centering
  \includegraphics[width=1.0\linewidth]{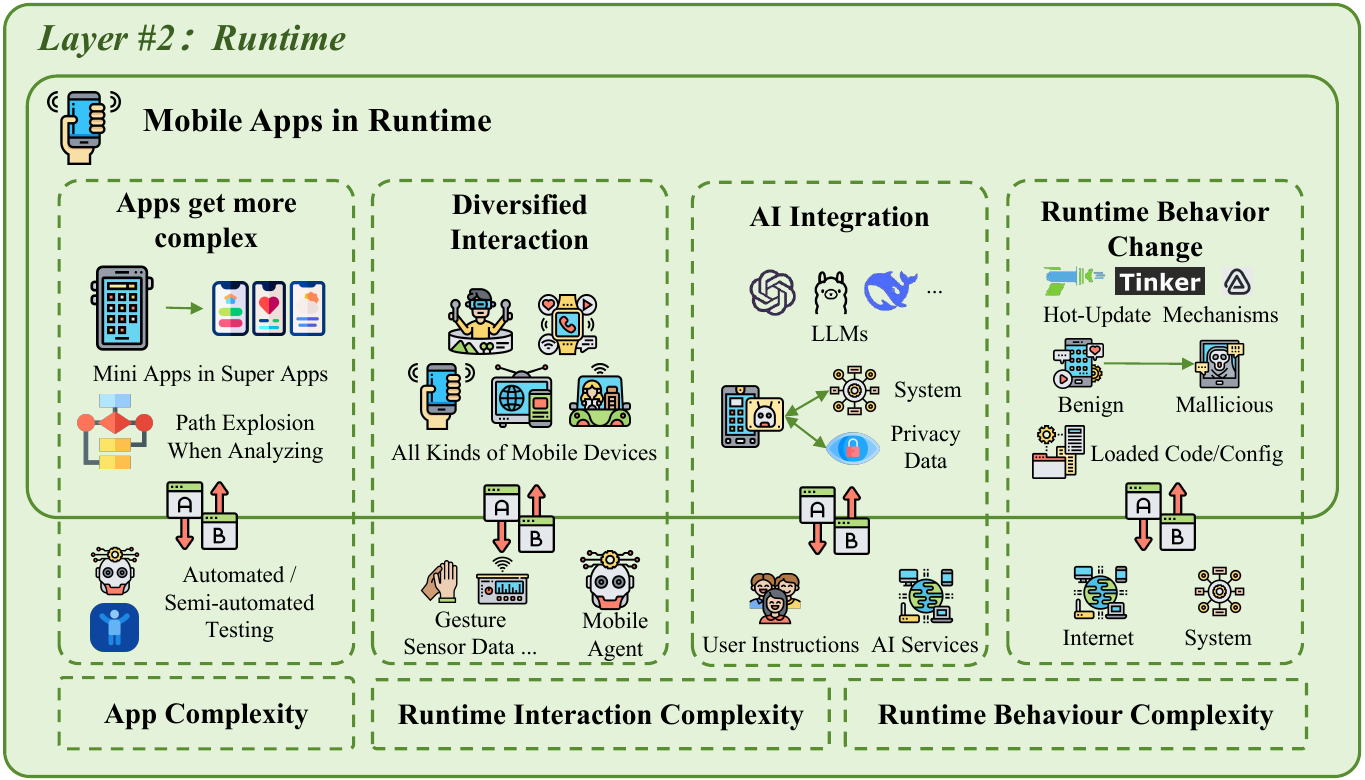}
  \caption{Overview of Runtime Layer}
  \label{fig:runtime-layer-fig}
\end{figure}

At the Runtime Layer, mobile apps exhibit three interconnected dimensions of complexity that reshape how analysis is expected to be conducted (as shown in \autoref{fig:runtime-layer-fig}). These characteristics reflect the evolution from simple, predictable execution patterns to dynamic, adaptive behaviors that challenge traditional analysis assumptions.

\textbf{App Complexity.} Modern mobile apps have evolved from simple, single-purpose tools into complex, multifunctional platforms. Many apps now integrate diverse services (messaging, payments, media streaming, social features) within single packages. The emergence of super-apps represents the extreme of this trend: platforms like WeChat combine messaging, payments, e-commerce, social networking, and mini-app ecosystems into unified apps~\cite{DBLP:journals/tem/SalehiMNG24}. These super-apps host millions of embedded mini-apps available for installation, with distinct functionalities, permission models, and interaction patterns~\cite{DBLP:conf/uss/WangZL23,DBLP:conf/ccs/Yang0L22}. Many super-apps support instant installation of mini-apps within the host environment, allowing users to access services without separate downloads. The architectural complexity covers cross-component data flows, nested execution contexts, and inter-app communication mechanisms that create dense, interconnected state spaces~\cite{DBLP:journals/tdsc/LiYYXTWZ24,DBLP:journals/tse/WangFLTJWXL24}.

\textbf{Runtime Interaction Complexity.} Beyond architectural complexity, modern mobile apps operate through diverse interaction paradigms that extend far beyond traditional touch-based interfaces, covering both the diversification of interaction modalities and the emergence of autonomous agents as new interaction subjects.

\begin{itemize}
    \item \textbf{Multi-Modal Interaction Paradigms.} Contemporary mobile interfaces support diverse interaction patterns beyond traditional touch input: gesture recognition, voice commands, augmented and virtual reality overlays, and cross-device coordination. These interaction modes are particularly prominent in emerging platform categories (AR/VR headsets, smartwatches, and automotive systems) where traditional touch-based interfaces are supplemented or replaced by spatial gestures, voice control, and contextual awareness. AR/VR interactions involve continuous spatial and temporal data streams, gesture recognition depends on timing and contextual factors, and cross-device scenarios involve coordination across multiple platforms simultaneously.

    \item \textbf{Agent-Driven Interaction.} Mobile apps are increasingly operated by LLM-driven agents~\cite{appagent,autodroid,mobileagent,mobileagentv2,autoglm} that act as intermediaries between users and apps. These agents accept natural language instructions from users, parse graphical interfaces to understand app states, and execute multi-step operations autonomously across apps and system services. Recent surveys highlight the growing importance of such computer-use agents~\cite{sager2025comprehensivesurveyagentscomputer}, while industry reports frame agent-driven interaction as an emerging mainstream channel for mobile scenarios~\cite{NEWSagent3}. The agent architecture combines semantic understanding through large language models, visual interpretation of GUI elements, and system-level operation execution. This introduces a dual nature: agents can serve as testing tools that automate complex interaction sequences, but their human-like operation patterns also expose them to risks. Traditional threats targeting human users (such as phishing interfaces or deceptive GUI elements) may affect agents due to their designed mimicry of human behavior. Additionally, agents face novel attack vectors specific to LLM systems, including prompt injection and environmental manipulation~\cite{DBLP:journals/corr/abs-2505-12981}.  \looseness=-1
\end{itemize}

\textbf{Runtime Behavior Complexity.} Complementing architectural and interaction complexity, app behavior is no longer static after installation. Apps now adapt their functionality at runtime through dynamic loading, hot-update, and AI integration, introducing dynamic and unpredictable behaviors that complicate analysis.

\begin{itemize}
    \item \textbf{Runtime App Behavior Change.} Modern apps can alter their behavior after deployment through hot-update frameworks such as CodePush~\cite{CodePush} for React Native, Shorebird~\cite{shorebird}, and Flutter Eval~\cite{fluttereval}. These enable over-the-air code delivery, bypassing app store review cycles. The mechanisms use hybrid execution engines to reconcile interpreted hotfix code with precompiled binaries, but this dual abstraction often introduces temporal inconsistencies and unpredictable state transitions. Although major distribution platforms restrict such practices~\cite{AppleBan,googleBan}, commercial solutions including Tencent Tinker~\cite{TinkerAndroidHotfix}, Alibaba AndFix~\cite{AndFixAndroidHotfix}, Hansel~\cite{HanselMobileJava}, and Apptimize~\cite{ApptimizeInstantUpdate} remain widely used. Dynamic code injection makes post-deployment behavior potentially different from pre-release analysis, undermining stability assumptions. Recent studies document that some market-listed apps exhibit discrepancies between storefront descriptions and after-installation behavior~\cite{Zhao24NoSource,Lee2019Understanding,Esposito2024Geofence,AndroidHeadlines2024Geofence}. These ``chameleon apps'' have been characterized in prior research~\cite{Zhao24NoSource}, showing how runtime changes can enable deceptive practices and complicate static analysis and regulatory oversight.\looseness=-1
  
    \item \textbf{AI Integration in Apps.} The direct embedding of generative AI capabilities represents a significant trend. A recent study \cite{hau2025llmsmobileappspractices} analyzed 149 open-source Android apps incorporating LLMs, showing that these models are primarily utilized as chatbots but also perform diverse functions such as specialized content generation, voice interaction, and task/productivity management. Integration approaches vary: some apps rely on third-party APIs from providers like OpenAI and Google, whereas others implement self-deployed APIs or embed lightweight models directly within the apps \cite{hau2025llmsmobileappspractices}.
    This integration paradigm enables sophisticated features including natural language interaction, intelligent agents, and autonomous decision-making components \cite{Hou24LLM4SESLR,chen2024llm}, delivering substantial efficiency and functionality. Contemporary mobile operating systems are increasingly incorporating LLMs as autonomous agents for automated tasks \cite{First24Wu,Understanding26Jin}.
    Also, LLM middleware frameworks (e.g., LangChain~\cite{langchain}) have also streamlined app creation by allowing developers to interface with LLMs through backend APIs, accelerating the growth of multifunctional apps. At the same time, this trend brings concrete risks: reliance on external APIs raises concerns over data leakage, on-device models demand heavy resources and careful isolation, and the nondeterministic nature of LLM outputs complicates testing and quality assurance.
\end{itemize}

These three dimensions of runtime complexity collectively transform the landscape for mobile app analysis. Unlike the relatively static challenges of the Artifact Layer, runtime complexity introduces temporal and contextual dependencies that make analysis inherently more difficult. The state space explosion from super-apps, the simulation challenges of multi-modal and agent-driven interactions, and the unpredictability introduced by hot-update and AI integration create a fundamentally different analytical context than what traditional dynamic analysis frameworks were designed to handle.

\subsection{Analysis Challenges}

The runtime characteristics outlined above create substantial challenges for mobile app analysis. These challenges affect different aspects of analysis but share a common theme: traditional approaches designed for simpler, more predictable apps struggle to handle modern runtime complexity.\looseness=-1

\textbf{Testing Complexity Challenges.} The first group of challenges concerns difficulties in testing modern apps comprehensively.

\begin{itemize}
\item \textbf{Coverage and Scalability.} Modern mobile apps generate state spaces that are difficult to explore exhaustively through conventional testing. The combinatorial explosion of execution paths across multiple subsystems, embedded apps, and cross-component data flows makes systematic coverage increasingly difficult~\cite{DBLP:journals/tdsc/LiYYXTWZ24,DBLP:journals/tse/WangFLTJWXL24}. Even recent advances in LLM-guided testing~\cite{DBLP:journals/pacmse/WangLZYW25} and automated event sequence generation~\cite{chen2025apexautomaticeventsequence} face difficulties as apps grow into multifunctional ecosystems. The result is that significant portions of app functionality remain unexplored, creating blind spots in security assessment.

\item \textbf{Interaction Simulation.} Multi-modal interactions involving gesture recognition, voice commands, AR/VR overlays, and cross-device coordination resist simulation in testing environments. These patterns are particularly problematic for emerging platform categories (AR/VR headsets, smartwatches, automotive systems) where apps rely on platform-specific sensors and interaction modes. These interaction patterns depend on continuous spatial and temporal data streams, nondeterministic timing, and contextual factors that are difficult to model and reproduce. Traditional GUI testing frameworks, designed for deterministic touch-based event sequences, have difficulty simulating the high-dimensional and context-sensitive behaviors required for comprehensive coverage. Testing environments struggle to accurately simulate gesture timing, spatial AR interactions, or cross-device synchronization patterns that depend on real-world contexts and hardware configurations.
\end{itemize}

\textbf{Autonomous Behavior Security Challenges.} The second group addresses security and privacy risks from intelligent, autonomous operations.

\begin{itemize}
\item \textbf{Agent-Driven Security.} As agents become mainstream interaction channels for mobile scenarios~\cite{NEWSagent3}, they expose risks already observed in non-mobile domains that become more acute in mobile contexts. Prompt injection and manipulation attacks can steer agents toward adversarial behaviors~\cite{liu2024promptinjectionattackllmintegrated,watchout}, backdoor and action hijacking threats can cause benign-looking tasks to trigger unintended privileged operations~\cite{zhang2024towards}, and environmental injection attacks exploit crafted interface elements or contextual cues to influence agent decision-making~\cite{liao2024eia,xu2024advwb,llmsthreat}. In mobile settings, these issues become more pressing because mobile ecosystems combine complex runtime states, permission systems, and cross-app deep link mechanisms that adversaries can exploit. Crafting deceptive GUIs, poisoned deep links, or hidden triggers can induce agents into granting dangerous permissions, conducting unauthorized purchases, or exfiltrating sensitive data. Recent studies confirm that mobile LLM agents can be exploited in practice~\cite{DBLP:journals/corr/abs-2505-12981}, meaning mobile app analysis needs to evaluate not only how apps might deceive human users but also how they can manipulate automated agents.\looseness=-1

\item \textbf{AI Integration Risks.} Embedding AI capabilities directly into apps creates privacy and security implications distinct from agent-driven risks. The diverse integration approaches (from third-party API dependencies to local model deployment~\cite{hau2025llmsmobileappspractices}) create varied attack vectors spanning prompt injection~\cite{liu2024promptinjectionattackllmintegrated}, model manipulation, and data exfiltration. The nondeterministic nature of LLM behaviors complicates reproducible testing and verification, as the same inputs may produce different outputs. Processing sensitive user data through AI components raises compliance challenges under regulations, particularly when AI decision-making processes are opaque. Traditional analysis methods struggle to capture semantic vulnerabilities in natural language interfaces or assess privacy implications when data flows through nondeterministic AI processing, creating gaps in both security assessment and regulatory compliance verification.
\end{itemize}

\textbf{Temporal Behavior Detection.} Apps can change their behavior after deployment through hot-update and dynamic loading. Hot-update mechanisms enable over-the-air code delivery that can alter app functionality, making pre-release analysis insufficient for understanding actual behavior. Although officially restricted by major platforms~\cite{AppleBan,googleBan}, commercial solutions remain adopted secretly. This enables ``chameleon apps'' that exhibit behavior discrepancies between storefront descriptions and installed functionality~\cite{Zhao24NoSource}. The temporal inconsistency undermines static analysis validity and creates challenges for detecting covert behavioral changes that occur after deployment. Current analysis frameworks generally lack the continuous monitoring capabilities needed to track post-deployment modifications.

These challenges show that runtime complexity has altered the requirements for mobile app analysis across multiple dimensions, from testing scale to interaction simulation to security assessment of autonomous behaviors to temporal behavior tracking.

\subsection{Potential Research Directions}

The challenges outlined above point toward research directions that address different aspects of runtime complexity.

\textbf{Adaptive Testing Frameworks.} The coverage and scalability challenges posed by super-apps and multifunctional apps demand testing approaches that move beyond exhaustive exploration toward intelligent, targeted analysis. Research should develop adaptive frameworks that use machine learning and large language models to guide state space exploration efficiently. These frameworks could incorporate several key capabilities: memory mechanisms that track exploration history to avoid redundant testing of equivalent states, learned models that predict which execution paths are more likely to show faults or security issues, and dynamic adjustment of testing strategies based on observed app behavior. Recent work on LLM-guided testing~\cite{DBLP:journals/pacmse/WangLZYW25,DBLP:conf/dtpi/LiLY24} and exploration with memory~\cite{zhang2025traceragllmbasedframeworkexplainable} provides foundations, but substantial work remains to handle the complexity of interconnected feature sets in modern apps. For multi-modal interactions, research should develop specialized simulation engines that can model and generate diverse interaction sequences, including gesture patterns with realistic timing variations, voice commands with different acoustic properties, and AR/VR spatial interactions that account for environmental contexts. The goal is semantic-aware testing that understands app functionality rather than merely exercising code paths.

\textbf{Security Analysis for Autonomous Operations.} The next two directions address security risks from intelligent behaviors, though they target different subjects and attack surfaces.

\begin{itemize}
\item \textbf{Agent Security Analysis.} Research should develop methods to systematically evaluate how apps might exploit or manipulate autonomous agents. This requires analyzing app interfaces for elements that could mislead agents through deceptive visual layouts, ambiguous labels, or hidden state indicators; detecting patterns that might trigger unintended agent actions through crafted deep links, permission requests, or interaction sequences; and verifying that apps do not include adversarial prompts or contextual cues designed to manipulate agent decision-making through prompt injection or environmental manipulation~\cite{DBLP:journals/corr/abs-2505-12981,liao2024eia,xu2024advwb}. Beyond defensive analysis, research should explore using agents as analysis tools themselves. Recent surveys on GUI agents with foundation models~\cite{DBLP:journals/corr/abs-2411-04890,DBLP:journals/corr/abs-2504-19838} suggest opportunities for agents to assist in automated testing, security assessment, and compliance verification. The challenge is developing bidirectional approaches where agents serve as both subjects requiring protection and tools providing analytical capabilities.

\item \textbf{AI Behavior Analysis.} Apps that integrate AI capabilities directly require methods to characterize the space of possible AI behaviors, bounding the range of outputs an AI component might produce for given inputs, identifying conditions under which AI behavior might become problematic, and detecting when AI responses might leak sensitive information or violate privacy expectations. This includes analyzing prompt injection vulnerabilities in mobile-specific contexts where AI components interact with system services and user data~\cite{liu2024promptinjectionattackllmintegrated}, tracking data flows through AI components to verify compliance with privacy regulations, and developing testing approaches that can provide meaningful coverage despite nondeterminism. The diverse integration patterns observed in practice~\cite{hau2025llmsmobileappspractices}, such as third-party APIs, local models, and middleware frameworks, require analysis techniques that can handle varied architectures while providing consistent security and privacy assessments.
\end{itemize}

\textbf{Comprehensive Runtime Behavior Monitoring.}  
The temporal gaps created by hot-update and dynamic code loading demand continuous monitoring techniques to detect post-deployment modifications. A central difficulty is the lack of systematic detection methods: existing approaches often rely on indirect indicators such as developer metadata or market descriptions~\cite{Zhao24NoSource}, but fail to reliably identify embedded frameworks or dynamic loading behaviors, leaving app marketplaces and analysis tools exposed to unmonitored changes. This problem is compounded by the divergence between installation and runtime behavior, where hot-update mechanisms allow apps to pass market review in one state but later alter functionality through OTA updates, enabling so-called “chameleon apps” that bypass pre-installation checks and pose severe security and regulatory challenges~\cite{Esposito2024Geofence}. Addressing these issues requires monitoring approaches that integrate static indicators (e.g., framework signatures, bytecode patterns) with dynamic runtime evidence, ideally within sandboxed environments that preserve execution integrity, so that covert hot-update mechanisms can be detected and the risks of unvetted code injection can be reduced.

These research directions address different aspects of runtime complexity: intelligent exploration for testing scale, security analysis for autonomous operations, and temporal monitoring for behavioral changes. Progress requires moving beyond traditional assumptions about predictable, static app behavior toward approaches that can handle the dynamic, adaptive, and intelligent characteristics of modern mobile apps.

\section{Ecosystem Layer}
\label{sec:ecosystem-layer}

\begin{figure}
  \centering
  \includegraphics[width=1.0\linewidth]{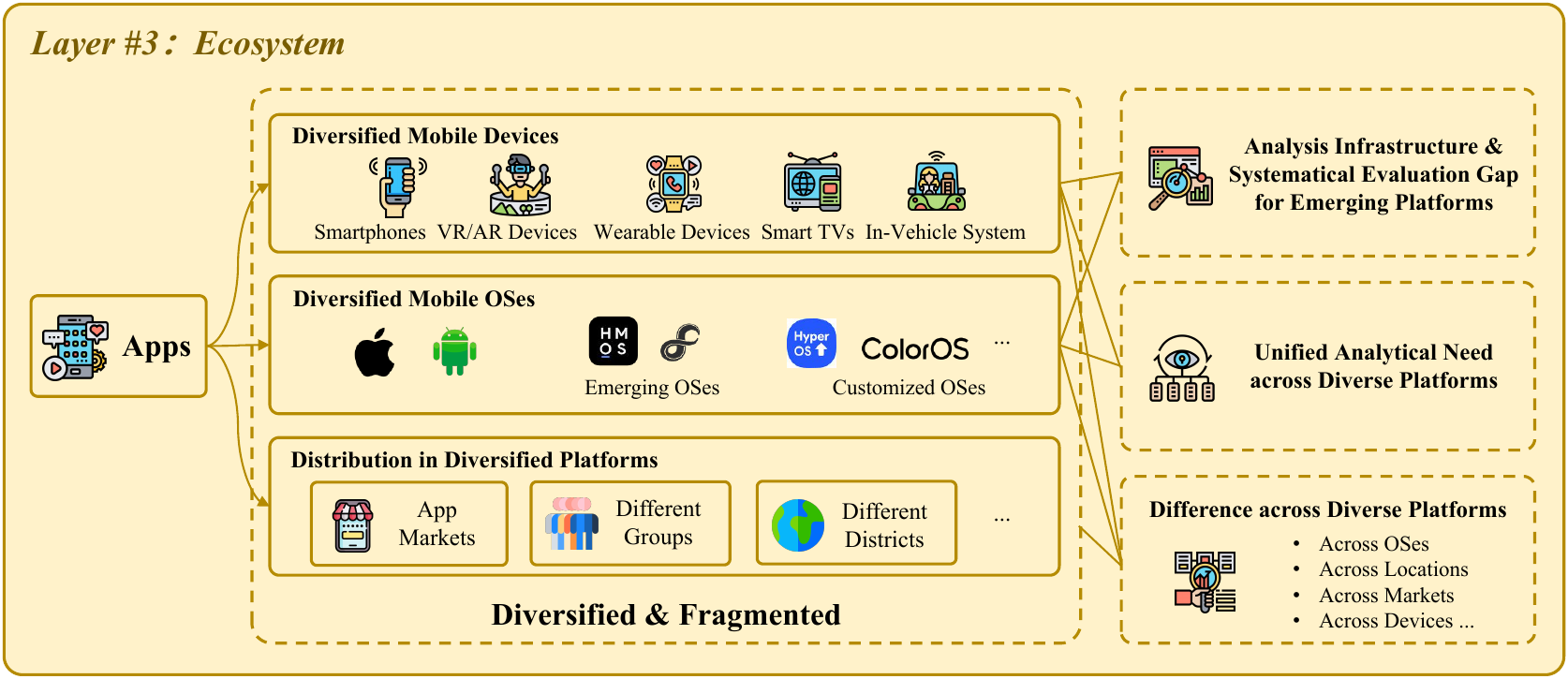}
  \caption{Overview of Ecosystem Layer}
  \label{fig:ecosystem-layer-fig}
\end{figure}

The Ecosystem Layer shifts the focus outward from apps themselves to the broader environments in which they are developed, distributed, and regulated. Unlike the Artifact and Runtime Layers, which examine internal structures and execution behaviors, this layer addresses how contextual diversity and distribution fragmentation reshape analysis requirements. Mobile apps now operate across an expanding range of devices, operating systems, and distribution channels. As illustrated in \autoref{fig:ecosystem-layer-fig}, these trends transform mobile app analysis from a uniform task into a context-dependent challenge that varies across platforms, markets, and regulatory contexts.

\subsection{Emerging Characteristics}

From the ecosystem perspective, mobile apps are no longer confined to a single dominant platform or device type. Instead, the landscape is both \emph{diversified}, with apps extending to new platforms, devices, and operating systems, and \emph{fragmented}, with the same app exhibiting differences across distribution channels, user groups, and geographic regions or other dimensions. This section highlights three key aspects of this evolution: diversified devices, diversified operating systems, and diversified distribution.

\textbf{Platform Diversification.} Mobile platforms have diversified in two complementary ways: through the proliferation of new device categories, and the rise of emerging operating systems.

\begin{itemize}
    \item \textbf{Diversified Mobile Devices.} Mobile platforms have extended beyond smartphones to AR/VR glasses, wearable devices, smart TVs, in-vehicle systems, and IoT hybrids, permeating daily life through apps that span health monitoring, spatial computing, smart home control, and immersive entertainment. These apps mirror smartphone functionalities while retaining platform-specific traits. Some apps operate on modified mobile operating systems: Meta Horizon OS (powering the Meta Quest series~\cite{metaquest}) and Android Automotive OS~\cite{AndroidAutomotive} leverage Android variants, while visionOS~\cite{visionos} builds on iPadOS foundations. This architectural inheritance highlights opportunities to transfer mobile app analysis techniques to related fields. \looseness=-1

    \item \textbf{Operating System Diversification.} Mobile apps traditionally targeted two dominant platforms, Android and iOS, but the operating system landscape has become more diverse. New mobile operating systems are emerging as viable alternatives: HarmonyOS~\cite{harmonyos} by Huawei reached approximately 19\% market share in China by end of 2024 and surpassed 5\% globally in early 2025, with over 1 billion active devices reported by Huawei~\cite{counterpoint_os_market_share}. Other systems have pursued different approaches but with limited impact: Tizen OS~\cite{TizenOS} and Ubuntu Touch~\cite{UbuntuTouch} saw modest mobile adoption and are no longer active in the smartphone market, while Google's Fuchsia~\cite{fuchsia} was deployed on smart-home devices but has since seen reduced maintenance. Android-compatible variants such as OPPO's ColorOS~\cite{ColorOS} and Xiaomi's HyperOS~\cite{HyperOS} modify core Android functionalities while maintaining app compatibility, but these modifications can affect how apps behave~\cite{Same24Dong}. Recent research shows that custom ROMs sometimes grant native privileges to pre-installed apps in ways that differ from stock Android~\cite{DBLP:conf/sp/WangZDW25}, altering the security assumptions that analysis tools rely on. The proliferation of both new operating systems and customized variants creates a landscape where apps exhibit platform-specific behaviors that require corresponding analysis adaptations.
\end{itemize}

\textbf{Distribution and Contextual Fragmentation.} Beyond platform diversity, mobile apps exhibit fragmentation across their deployment contexts, creating variations in how the same app operates in different settings. Apps are distributed through multiple channels like official app stores like Google Play and Apple App Store, regional stores like Tencent App Store~\cite{TencentAppstore}, Huawei AppGallery~\cite{HUAWEIAppGallery} and even via Telegram~\cite{DBLP:journals/pomacs/GuoWWFWYLW24} or other underground distribution platforms. Each distribution channel applies different review processes, update schedules, and content requirements. Geographic markets introduce additional variation: apps adapt their behavior based on deployment region due to regulatory requirements, infrastructure differences, and local market conditions. Recent studies document significant geodifferences in mobile apps~\cite{DBLP:conf/uss/KumarVR0E22} and substantial regional variations in security and privacy features~\cite{conf/sp/GuoNLC25}. Apps also modify their interfaces, available features, and data handling based on user characteristics including age, language preferences, and accessibility needs. Age verification mechanisms vary significantly across regions and contexts~\cite{10.5555/3766078.3766080}, while privacy practices adapt to different user populations~\cite{DBLP:journals/popets/ZimmeckSSRWRRS19,DBLP:journals/popets/MoremenHB24}. This fragmentation means apps can exhibit different code and behaviors depending on their specific deployment context, making uniform analysis insufficient for understanding actual app behavior across markets and user groups.

These trends transform mobile app analysis from a uniform task into a context-dependent challenge that account for platform-specific constraints, regulatory requirements, and market variations while maintaining analytical consistency across heterogeneous environments.

\subsection{Analysis Challenges}

Platform diversification and distribution fragmentation introduce challenges that manifest in different parts of the analysis process. We summarize them into four categories.

\textbf{Emerging Platform Capability Gaps.}  
New operating systems and vendor-customized ROMs often lack mature infrastructures for analysis. Custom ROMs illustrate this issue: while they preserve app compatibility, they also modify core system behaviors in ways that undermine existing assumptions. Recent studies show that some ROMs grant elevated privileges to pre-installed apps, breaking Android’s permission model~\cite{DBLP:conf/sp/WangZDW25}. HarmonyOS poses a broader challenge: its ArkTS language~\cite{arkts_language} and distinct execution model invalidate Dalvik/ART-based tools, creating an analytical void. Although initial efforts such as Ark-Analyzer~\cite{chen2025arkanalyzer} provide partial support, comprehensive static and dynamic pipelines are still missing. Dynamic testing frameworks also struggle, as instrumentation tools need to be redesigned for each new system variant. Beyond individual platforms, systematic app-level analysis remains sparse: while some work has addressed VR apps~\cite{EmpiricalOculus} and Android TV apps~\cite{LiuAndroidTV}, most research remains at the system or device level~\cite{KhanWearable,rathore2022vehicle,LiuSmartwatch}, leaving app-centric risks underexplored.

\textbf{Policy and Compliance Challenges.}  
Diverse platforms impose heterogeneous policies that complicate verification. Android’s Car App Quality guidelines~\cite{AndroidCarAppQuality} limit video, gaming, and browsing functionality in driving mode, while Wear OS standards~\cite{AndroidWearOSAppQuality} constrain smartwatch UIs by forbidding certain elements. These rules are highly context-sensitive and not easily captured by existing analysis workflows. Moreover, compliance requirements extend beyond platform rules to jurisdiction-specific regulations such as GDPR~\cite{GDPR2024}, requiring app behavior to be validated simultaneously against multiple, potentially conflicting regimes.

\textbf{Technical Model Diversity Challenges.}  
Underlying execution differences add another layer of complexity. HarmonyOS replaces Android’s runtime with ArkTS, introducing new abstractions. Custom ROMs alter privilege granting and permission handling. Wearables and AR/VR platforms integrate sensor-rich, context-dependent interfaces that differ fundamentally from smartphones. Current analysis tools, which assume a stable execution model and permission system, lack the flexibility to reason consistently across these heterogeneous environments.

\textbf{Cross-Context Systematic Comparison Challenges.}  
Distribution fragmentation leads to variations in app behavior across markets, stores, and user groups. Large-scale studies show significant geodifferences in functionality~\cite{DBLP:conf/uss/KumarVR0E22} and substantial regional disparities in privacy and security~\cite{conf/sp/GuoNLC25}. Age verification practices differ across jurisdictions and can often be circumvented via VPNs or OAuth misuse~\cite{10.5555/3766078.3766080}. Privacy compliance analysis at scale~\cite{DBLP:journals/popets/ZimmeckSSRWRRS19} and generational differences in privacy attitudes~\cite{DBLP:journals/popets/MoremenHB24} further demonstrate how contextual factors shape app behavior. Current frameworks lack the ability to systematically compare across such variations, making it difficult to ensure consistent compliance and detect region-specific risks.

These challenges show that ecosystem-level complexity requires analysis approaches that can handle infrastructure diversity, policy heterogeneity, and contextual fragmentation while providing systematic coverage across the expanding mobile landscape.

\subsection{Potential Research Directions}

The challenges posed by ecosystem diversification and fragmentation point toward research directions that address complementary aspects of the problem. Some directions focus on building analytical capabilities for emerging platforms, while others aim to provide systematic ways of ensuring consistency across heterogeneous environments.

\textbf{Building Platform-Specific Capabilities.}  
Emerging operating systems and customized variants require dedicated infrastructures to close current capability gaps.  
\begin{itemize}
  \item \textbf{Platform-Specific Infrastructure Development.} Develop static and dynamic pipelines that natively support new languages and runtimes, such as ArkTS for HarmonyOS~\cite{arkts_language}, beyond Dalvik/ART-based approaches. This includes instrumentation frameworks and automated testing tailored to new execution semantics. While early tools like Ark-Analyzer~\cite{chen2025arkanalyzer} provide a starting point, more complete toolchains approaching the maturity of Android and iOS ecosystems are still needed.  
  \item \textbf{Legacy Technique Adaptation.} Established methods such as taint analysis and GUI testing should be systematically adapted. Proven approaches remain valuable if adjusted to platform-specific permission models and interaction paradigms, helping to bridge mature knowledge with emerging needs.  
\end{itemize}

\textbf{Policy-Aware Verification.}  
The heterogeneous compliance landscape across platforms and devices suggests the need for automated mechanisms that can account for diverse policy requirements. One promising direction is to develop policy-as-code frameworks that encode platform-specific rules, such as Android Car App Quality standards~\cite{AndroidCarAppQuality} or Wear OS App Quality guidelines~\cite{AndroidWearOSAppQuality}, as verifiable constraints integrated into analysis workflows. At the same time, these systems could be extended to incorporate jurisdiction-specific regulations (e.g., GDPR~\cite{GDPR2024}), moving toward automated assessments of app behavior across fragmented deployments against both platform-level guidelines and regional laws.

\textbf{Toward Unified Cross-Platform Analysis.}  
The proliferation of diverse platforms calls for analytical approaches that can ensure consistent coverage while remaining adaptable to platform-specific constraints. On one hand, apps on emerging platforms such as wearables, AR/VR, and automotive systems expose unique context-sensitive risks, requiring scenario-specific simulation in dynamic testing environments (as discussed in \S\ref{sec:runtime-layer}). On the other hand, many of these apps share implementation logic through cross-platform development frameworks, which encapsulate common code bases across Android, iOS, and other platforms (see \autoref{fig:artifact-layer-fig}). This shared nature provides opportunities for unified analysis that leverages commonalities rather than treating each platform in isolation. Yet current practice remains fragmented: existing tools still target platforms separately, and well-established frameworks for joint analysis are lacking. Bridging this gap requires methods that can both exploit shared implementations for efficiency and incorporate platform-specific adaptations for completeness. Developing such unified approaches represents a central challenge for mobile app analysis in the multi-platform ecosystem.

\textbf{Ecosystem-Scale Comparative Analysis.}  
The fragmentation of app distribution across stores, regions, and demographic contexts highlights the need for systematic frameworks that can detect and analyze behavioral variations across different deployment scenarios. Automated comparison systems could help identify discrepancies in features, configurations, and practices across markets, languages, and user groups, building on recent studies of geodifferences~\cite{DBLP:conf/uss/KumarVR0E22} and regional security variations~\cite{conf/sp/GuoNLC25}. Incorporating compliance checks with legal regimes such as GDPR~\cite{GDPR2024} and enabling longitudinal tracking of app evolution across platforms and jurisdictions would further strengthen the analytical toolkit.

Together, these research directions combine platform-specific adaptation with broader unification and systematic comparison. Progress in this area will likely depend on balancing efforts to address immediate capability gaps for new platforms with the development of generalizable principles that maintain analytical relevance across an increasingly diverse mobile ecosystem.

\section{Discussion}
\label{sec:discussion}

The preceding sections have examined challenges at the Artifact, Runtime, and Ecosystem Layers. To bring together observations that were relatively scattered across these sections, we focus on two specific topics here. First, \S\ref{subsec:AIDiscussion} provides a more in-depth discussion of how large language models, currently a prominent technological trend, affect mobile app analysis. Then, \S\ref{subsec:MultiDimensionDiscussion} discusses the overall development trend of Mobile App Analysis from a multi-dimensional perspective (analytical capabilities, analytical needs, and stakeholders), aiming to provide some potentially valuable insights.  \looseness=-1

\subsection{LLMs: Opportunities and Challenges for Mobile App Analysis}
\label{subsec:AIDiscussion}

Among all the transformations discussed in this paper, large language models (LLMs) and related techniques (particularly intelligent agents powered by LLMs), stand out as qualitatively new factors for mobile app analysis. LLMs and agents not only enable new analytical capabilities but also introduce novel security challenges. In this subsection, we discuss how LLMs and agents are reshaping the analysis pipeline, what new capabilities and constraints they introduce, and what promising directions emerge for advancing mobile app analysis.

\subsubsection{LLM as Enabler.}\

\noindent \textbf{LLM-Assisted Dynamic Analysis.} Dynamic analysis has emerged as a primary domain where LLMs are being integrated into mobile app analysis. In these approaches, LLMs serve as intelligent decision-makers that can replace human judgment in testing workflows, bringing semantic understanding to tasks that traditionally required manual expertise.

LLMs are used to guide exploration of app state spaces by making context-aware decisions about which actions to execute and which states to explore. Comprehensive testing approaches like GPTDroid~\cite{liu2024make} and LLMDroid~\cite{DBLP:journals/pacmse/WangLZYW25} employ LLMs for functionality-aware navigation throughout testing sessions, aiming to achieve higher testing coverage. Intent-driven and scenario-driven testing methods~\cite{droidagentgit,beyzaei2025automated,zhang2024llm} leverage LLMs to understand testing objectives and guide exploration accordingly. Advanced approaches incorporate memory mechanisms~\cite{zhang2025traceragllmbasedframeworkexplainable} to maintain context across testing sessions, enabling more sophisticated behavioral analysis that adapts based on observed app behavior.
Beyond navigation, LLMs are employed to generate and adapt testing inputs by understanding the semantic requirements of app interfaces. Test migration approaches~\cite{cao2025intention,beyzaei2025automated,zhang2024llm} employ LLMs to adapt test scripts across functionally similar apps by recognizing semantic equivalences between different implementations. Automated test input generation~\cite{cui2024large,QTypist,liu2024testing} addresses the challenge of generating valid contextual inputs by using LLMs' semantic understanding of UI components and their expected input formats. Bug reproduction tools~\cite{Feng2024AdbGPT,wang2024feedback,Icse2025crashtranslator} leverage LLMs to interpret crash reports and generate reproduction steps, replacing manual debugging analysis.

LLMs are also applied to assess app behavior quality by interpreting runtime observations against expected functionality. Non-crash functional bug detection~\cite{ju2024study,liu2024vision} demonstrates how LLMs can identify subtle malfunctions that require human-level understanding of expected app behavior, addressing cases where traditional test oracles are insufficient. Accessibility and usability issue extraction during testing~\cite{conf/icse/HuqTKM25} further illustrates how LLMs perform quality assessment tasks that traditionally required manual inspection.

These applications demonstrate how LLMs bring human-like reasoning to automated testing workflows. By supporting test navigation, input generation, and behavior evaluation, LLMs are enabling more intelligent and adaptive analysis that can handle modern app complexity. As these techniques mature, we anticipate further improvements in test coverage, bug detection accuracy, and the ability to handle increasingly complex app behaviors.

\noindent \textbf{LLM-Assisted Static Analysis.} While non-mobile domains have extensively adopted LLMs for static analysis, including reverse engineering~\cite{hu2024degpt,tan2024llm4decompile}, vulnerability detection~\cite{conf/uss/Liu0LDCYYSLZ0025}, and even program analysis assistance such as call graph construction~\cite{cgbench2025}, mobile-specific applications remain limited but show promising research potential.

Current mobile static analysis research demonstrates several emerging directions. Bytecode-level summarization and dataflow graph generation~\cite{androbyte2025} uses LLMs to bridge the semantic gap between low-level bytecode and high-level behavioral understanding. Code summarization for specific purposes includes malicious behavior detection~\cite{zhao2025apppoet,li2024enhancing} and secret leakage analysis~\cite{llmsecrets2025}, where LLMs interpret code semantics to identify security-relevant patterns. LLM-assisted deobfuscation~\cite{DBLP:journals/tifs/ZhaoZKPDZX25} represents another promising area where LLMs help recover semantic information from protected applications. These methods demonstrate the potential of LLMs for mobile-specific program analysis, but their reliability and interpretability in security-critical settings remain open questions that warrant further investigation.

\noindent \textbf{LLM-Driven Direct Analysis of App-Associated Data.}
Beyond assisting traditional static and dynamic analysis, an emerging direction leverages LLMs to directly analyze app code and associated artifacts, such as privacy policies, app descriptions, user reviews, and documentation, without relying on static analysis frameworks or runtime execution. By harnessing their semantic understanding capabilities, LLMs can process these diverse data sources in ways that traditional analysis tools cannot, opening complementary perspectives on app behavior that go beyond code-level inspection. \looseness=-1

LLMs are now used to analyze app-associated textual data including privacy policies, app descriptions, and user-generated content that traditional tools cannot thoroughly understand and summarize. Privacy policy analysis~\cite{DBLP:conf/uss/CuiTMJ23} and sensitive behavior description synthesis~\cite{appbds2025} extract privacy implications from natural language documents to verify policy-behavior consistency and compliance. These approaches enable automated assessment of what apps claim to do versus what they actually implement.

LLMs can also enable ecosystem-level studies when applied across large app corpora or extensive user-generated content. Large-scale analysis of user reviews~\cite{Akgul24decade} extracts privacy-related concerns and trends from thousands of apps, synthesizing user perspectives that would be impractical to analyze manually. Android documentation analysis~\cite{Yan24Documented} processes platform specifications to extract privacy policy-related information at scale. Regional app variant comparison~\cite{conf/sp/GuoNLC25} uses LLMs to summarize functional differences across geodistributed versions of the same apps. The distinguishing characteristic of these studies is their scale and heterogeneity: LLMs synthesize insights from diverse data sources to identify patterns, inconsistencies, and emerging trends across the app ecosystem that would remain invisible in single-app analysis.

LLMs also facilitate the translation between program behavior and natural language, making technical analysis results accessible for compliance verification and transparency purposes. Behavior-policy consistency checking~\cite{DBLP:conf/ndss/Yue0GD0Y25} employs LLMs to generate natural language behavioral descriptions from static and dynamic analysis results for verification against stated policies. AI capability identification~\cite{llmaid2025} and sensitive data usage intention analysis~\cite{deng2025walls} demonstrate how LLMs interpret complex app behaviors for specific analytical objectives, converting technical findings into human-understandable explanations suitable for regulatory compliance and user transparency.

As LLM capabilities mature, we expect this direct analysis approach to play an increasingly important role in app compliance verification and ecosystem-level understanding, complementing traditional code-based analysis methods.

\noindent \textbf{LLM-Powered Agent for Autonomous Analysis.}
Beyond utilizing LLM capabilities within traditional analysis workflows, LLM-powered intelligent agents represent an emerging direction for mobile app analysis. Agents possess human-like reasoning capabilities and, empowered by mechanisms like function tool integration, memory management, and iterative planning, can accomplish complex automated tasks that simple LLM invocations cannot achieve~\cite{llm_agents_se_survey_2024}. Software engineering research begins exploring agent capabilities for code analysis tasks. Agents have shown promise in vulnerability detection~\cite{llm_vulnerability_survey_2025,jitvul_2025,av_agent_2025}, reverse engineering~\cite{ai_reverse_engineering_survey_2025,llm_malware_analysis_2025} and other fields highly related to software analysis. For mobile app analysis, agents' autonomous capabilities hold potential for extending beyond LLM-assisted methods. In dynamic testing, rather than invoking LLMs to guide individual test actions within fixed frameworks, agents can autonomously orchestrate testing campaigns, deciding when to switch strategies, adjusting plans based on intermediate results, and coordinating multiple analysis phases. Multi-agent frameworks~\cite{madroid_2025,mobile_agent_e_2025,xuat_copilot_2024} demonstrate how specialized agents can decompose complex testing tasks and coordinate their execution, handling scenarios that require adaptive synchronization across multiple components. Approaches that leverage retrieval mechanisms~\cite{cat_wechat_2024} further suggest how agents could efficiently reuse analysis experiences across similar apps, reducing redundant effort while maintaining effectiveness. 

For static analysis and reverse engineering, agents' ability to actively select code segments for examination, annotate findings, and maintain analysis context through memory mechanisms offers new possibilities for handling mobile apps' inherent complexity. Recent work in non-mobile domains demonstrates agents' effectiveness in vulnerability detection~\cite{jitvul_2025,av_agent_2025}, malware analysis~\cite{llm_malware_analysis_2025}, and automated code annotation~\cite{ai_reverse_engineering_survey_2025}. Mobile apps present additional complexity: extensive interactions among components like UI widgets, callbacks, and communication channels—interactions that traditional work carefully models~\cite{li2015iccta,DBLP:conf/icse/ChenTPL25}. Agents with memory could potentially maintain higher-level understanding of these global interactions, going beyond component-level analysis. While these foundational works provide promising directions, adapting these approaches to mobile-specific contexts is still in early stages.

Beyond these specific applications, agents' most distinctive capability lies in orchestrating existing mobile app analysis infrastructure. Rather than replacing individual analysis components, agents could autonomously plan and coordinate multiple tools—combining static analyzers, dynamic monitors, deobfuscation utilities, and semantic analysis—based on analysis objectives and intermediate findings. For instance, an agent analyzing privacy practices might: examine the manifest to identify relevant components, invoke taint analysis for specific flows, trigger dynamic testing to validate behaviors, cross-reference with policy text, and synthesize findings into structured reports. This autonomous tool orchestration represents a fundamentally different paradigm from embedding LLMs within individual analysis tools. Realizing this potential, however, requires addressing key challenges in agent architecture design, multi-tool coordination, and reliability guarantees. As agent technologies mature, we anticipate their integration with mobile app analysis infrastructure will enable more sophisticated and adaptive approaches, though ensuring reliability and controllability remains an open research question.

\noindent \textbf{Overall Trends and Future Directions.} 
These applications show consistent patterns in how LLMs are being integrated into mobile app analysis. LLMs primarily serve as components that bring semantic understanding to tasks traditionally requiring human expertise, such as interpreting user intent, evaluating behavior correctness, synthesizing insights from textual data, and, as agents, even orchestrating existing tools for intelligent analysis workflows. They are enabling analysis at substantial scales, from automating repetitive testing workflows to extracting patterns across entire app ecosystems.

Realizing the full potential of LLMs and agents in mobile app analysis requires addressing fundamental challenges. Behavioral nondeterminism and relatively high computational costs may limit their widespread adoption in production environments. How to enable more efficient, controllable, and cost-effective use of LLMs remains a critical research question. Future work should focus on developing practical frameworks that balance the power of LLMs with the reliability and cost-effectiveness required for industrial deployment. We anticipate that as these challenges are addressed, LLM-based approaches will become increasingly integral to mobile app analysis workflows, complementing rather than replacing traditional techniques.

\subsubsection{LLM as Challenge.}\

While LLMs offer powerful capabilities for mobile app analysis, their integration into mobile applications simultaneously introduces new security and privacy risks that create fresh analytical challenges. These challenges manifest in two primary dimensions: the security and privacy implications of embedding LLM functionality within mobile apps, and the novel attack surfaces created when LLM-driven agents operate mobile applications autonomously.

\noindent \textbf{LLM Integration Privacy and Security Challenges.}
Mobile apps increasingly embed LLM-backed features for content generation, recommendation, and conversational interaction~\cite{hau2025llmsmobileappspractices}. This integration introduces vulnerabilities that traditional mobile app analysis techniques were not designed to detect. Research on non-mobile LLM applications has established the threat landscape: prompt injection attacks enable unauthorized access and content manipulation~\cite{liu2023prompt_injection_houyi,greshake2023indirect_prompt_injection}, jailbreak techniques bypass safety mechanisms~\cite{chao2024jailbreakbench}, environment manipulation exploits crafted content~\cite{liao2024eia,xu2024advwb}, and privacy leakage occurs through training data extraction or contextual inference~\cite{carlini2021extracting_training_data,kim2024llm_privacy_leakage}. Ecosystem-scale analysis shows widespread violations, with thousands of LLM-integrated applications collecting sensitive information against stated privacy policies~\cite{hou2024llm_app_store_security}.

Mobile environments amplify these risks due to the concentration of sensitive data, for instance, location history, biometric data, health records, financial credentials, and continuous sensor streams~\cite{nowsecure2025owasp_mobile}. When mobile apps integrate LLM capabilities, user data may be processed through third-party API services or on-device models, creating new pathways for potential data leakage. Mobile-specific attack vectors compound these concerns: prompt injection through notification channels~\cite{chen2025aeia}, manipulation via crafted GUI elements~\cite{DBLP:journals/corr/abs-2505-12981}, and accessibility feature exploitation~\cite{nowsecure2025owasp_mobile} can lead to unauthorized access or unintended data disclosure. 

From an analysis perspective, these developments create fundamental challenges. Traditional mobile app analysis tools designed for deterministic logic face difficulties in tracking data flows through nondeterministic AI components, where the same input may produce varied outputs. Assessing privacy compliance becomes more complex when data processing involves opaque model inference, making it difficult to verify whether apps handle user data as claimed in privacy policies. Connecting observable AI behavior to specific code locations for vulnerability assessment requires new analytical approaches that bridge the gap between model-level operations and implementation-level code. While supply-chain defenses such as content filtering and pre-execution validation can mitigate certain attack vectors~\cite{wu2024prompt_defense,chen2025secalign}, developing systematic analysis capabilities for verifying privacy compliance and detecting security vulnerabilities in AI-integrated mobile applications remains an important research direction as LLMs become more deeply embedded in mobile ecosystems. \looseness=-1

\noindent \textbf{Mobile Agent Attack Surfaces.}
LLM-driven agents that operate mobile applications autonomously \cite{wang2024agent_survey} introduce attack surfaces where malicious apps exploit agent mechanisms. Non-mobile agent research has identified fundamental vulnerabilities: adversaries can manipulate agents through hidden prompts in content~\cite{greshake2023indirect_prompt_injection}, malicious instructions via notifications~\cite{chen2025aeia}, and jailbreak techniques including role-playing and adversarial suffixes~\cite{shen2023dan,zou2023gcg}. These attacks exploit the control-data confusion inherent in LLMs, where instructions and user content are processed through the same pathway.\looseness=-1

In mobile environments, these vulnerabilities carry amplified consequences due to broader system access and pervasive device usage. Empirical evaluations show alarming success rates: touch-guided jailbreak attacks achieved over 80\% success against state-of-the-art mobile agents~\cite{touch_jailbreak2024}. Attackers exploit mobile-specific vectors including crafted UI elements with invisible text, poisoned deep links, and manipulated screenshots~\cite{DBLP:journals/corr/abs-2505-12981,touch_jailbreak2024} to induce agents to grant dangerous permissions, execute unauthorized operations, or perform cross-application attacks~\cite{DBLP:journals/corr/abs-2505-12981}. The systemic challenge arises from agents' dual vulnerability: their human-like operation patterns make them susceptible to traditional interface-based deception (phishing UIs, deceptive dialogs), while their LLM foundation exposes them to model-specific attacks like prompt injection and reasoning manipulation~\cite{DBLP:journals/corr/abs-2505-12981,wang2024agent_survey}.

This requires extending mobile app analysis beyond evaluating whether apps might deceive human users to assessing whether apps might manipulate autonomous agents. Systematic investigation of mobile agent attack methodologies, development of analysis techniques to identify agent-exploitable vulnerabilities in apps, and establishment of security standards for agent-enabled mobile applications remain critically underdeveloped as mobile operating systems increasingly integrate agent capabilities (e.g., Apple Intelligence~\cite{apple_intelligence_2024}, Google Gemini~\cite{google_gemini_assistant_2025}).

\subsection{Overall Implications of Mobile App Analysis Evolution}
\label{subsec:MultiDimensionDiscussion}

The challenges identified in \S\ref{sec:artifact-layer}--\S\ref{sec:ecosystem-layer} affect mobile app analysis in different ways. To move from observing these challenges to understanding their implications, this section examines how analysis capabilities have changed, how requirements have evolved, and what stakeholders can do in response. This synthesis clarifies the practical significance of the transformations discussed earlier.

\subsubsection{Changes in Analysis Capabilities}\

Mobile app evolution creates substantial challenges for analysis, as discussed in previous sections. A critical issue is that code transparency faces increasing obstacles. Protection mechanisms, cross-platform frameworks, and multi-language compositions reduce what can be directly observed in packaged artifacts (\S\ref{sec:artifact-layer}). Runtime complexity further complicates analysis, AI components introduce nondeterministic behaviors and raise security and privacy concerns that lack systematic analytical approaches, while dynamic update mechanisms enable post-deployment behavior changes (\S\ref{sec:runtime-layer}).

At the same time, new analytical capabilities are emerging. LLM-based approaches bring semantic understanding to tasks that previously required manual effort or remained infeasible: analyzing obfuscated code, interpreting privacy policies, and connecting implementation behavior to policy descriptions for more direct verification (\S\ref{subsec:AIDiscussion}). Infrastructure improvements provide better foundations. Modern frameworks like Tai-e~\cite{tan2023taie} offer more scalable, efficient, and accurate program analysis capabilities for mobile app analysis~\cite{DBLP:conf/icse/ChenTPL25}, while industrial tools such as Fastbot~\cite{fastbot2,fastbot2_github} for automated GUI testing and AppShark~\cite{Appshark} for static analysis continue to mature.

\subsubsection{Evolution of Analysis Requirements}\

The capability changes discussed above directly influence what analysis is expected to achieve. As traditional approaches face limitations, new scenarios emerge that demand different analytical strategies, while the scope of what needs to be analyzed expands beyond individual apps.

Regarding scenarios, requirements have diversified beyond traditional code verification. Agent-based interaction introduces risks where apps may manipulate autonomous agents rather than human users. Dynamic update mechanisms create needs for continuous monitoring rather than one-time pre-release assessment. Environmental awareness and context-dependent behaviors—such as detecting malicious actions triggered only in specific conditions—require runtime monitoring capabilities. Multi-modal interactions through voice, gesture, and AR/VR interfaces require simulation capabilities that traditional testing lacks. AI integration into mobile apps raises questions about behavioral nondeterminism and privacy implications, though systematic analytical approaches for these issues remain immature. These scenarios demand techniques tailored to their specific characteristics.\looseness=-1

Regarding scope, traditional mobile app analysis focuses on individual applications in order to verify security properties, detecting vulnerabilities, and ensuring compliance within single apps. While this remains the core requirement, ecosystem-level analysis has become relevant: apps may behave differently across distribution channels, geographic markets, and regulatory jurisdictions. Apps providing similar functionality may exhibit different behaviors on diverse mobile device types, and analyzing these variations requires platform-specific infrastructure for systematic assessment. Effective analysis thus requires considering both individual app behavior and its ecosystem context.

\subsubsection{Implications for Stakeholders}\

The gaps between current capabilities and evolving requirements create challenges for different participants in the mobile ecosystem. Key stakeholders directly involved in mobile app analysis include developers who build and test apps, app markets and regulators who review and govern apps, and researchers who develop analysis techniques and infrastructure. Throughout previous sections, we have discussed various analysis challenges and emerging approaches from researchers' perspectives. Here, we examine how these changes affect developers and markets/regulators, and what actions they can take in response.

\noindent\textbf{Developers and App Vendors.}
Developers primarily use mobile app analysis for quality assurance during development and testing phases. They face increasing complexity in ensuring app quality as modern development introduces diverse technical components, each requiring specialized testing approaches. AI functionality integration further complicates quality assurance: nondeterministic model outputs make exhaustive state coverage difficult, and traditional testing tools lack support for these emerging components. Current toolchains provide only partial coverage, as new frameworks and AI integrations outpace the development of corresponding testing infrastructure.

However, LLM-assisted development and testing workflows offer promising opportunities to address some of these gaps. As discussed in \S\ref{subsec:AIDiscussion}, LLM-guided exploration can intelligently navigate complex state spaces, agent-based testing can simulate sophisticated user interactions, and automated test generation can reduce manual effort. Developers who engage with these emerging techniques can benefit from improved testing efficiency, though establishing robust quality assurance processes for AI-integrated apps remains an evolving practice.

\noindent\textbf{App Markets and Regulators.}
App Markets often use mobile app analysis to conduct security and privacy compliance reviews of apps before they are distributed. They face growing difficulties in pre-deployment review. As discussed in \S\ref{sec:artifact-layer}, the transparency of packaged artifacts is strongly reduced. Runtime behavior changes (\S\ref{sec:runtime-layer}) add further complications, as apps may behave differently after passing initial review. The same app may also exhibit inconsistent practices across geographic markets and distribution channels.

Therefore, proactive measures should be taken to address these difficulties. This involves actively researching and adopting adaptive analysis techniques that can handle code protection. On the other hand, given the difficulty of directly analyzing the application's code implementation, it is also worthwhile to actively explore alternative methods. For instance, as indicated in \S\ref{sec:runtime-layer}, exploring continuous and systematic dynamic monitoring that tracks post-deployment behavior becomes increasingly important. Additionally, analyzing metadata beyond code itself, or combining such metadata with partially extractable features from protected code, has become more valuable with the emergence of LLMs, as discussed in \S\ref{subsec:AIDiscussion}. These complementary approaches deserve continued exploration. \looseness=-1

Areas where analytical capabilities fall short also often present policy gaps. For example, while runtime updates are prohibited and related SDKs are blacklisted, developers can still tamper with their apps and update applications using flexible frameworks or even their own custom-designed techniques. Therefore, it is crucial to combine effective analytical techniques with clear rules to quantify such tampering, thereby protecting user rights and maintaining the order of the application ecosystem. Meanwhile, existing regulatory frameworks do not yet fully address emerging technologies like autonomous agents and embedded AI. Regulators may need to update existing frameworks to address emerging technologies.

\section{Conclusion}
\label{sec:conclusion}

By examining mobile app analysis through a three-layer perspective, this paper has provided a structured way to understand how modern apps challenge established methods. We have outlined how heterogeneous artifacts, increasingly complex runtime behaviors, and ecosystem-level fragmentation compound across layers, creating difficulties that traditional static and dynamic approaches struggle to address.  

To respond, we have proposed a layered framework that not only structures the space of challenges but also highlights recurring themes: the loss of analyzability in packaged artifacts, the explosion of runtime complexity, and the difficulty of maintaining consistency across fragmented platforms and regulatory contexts. Building on this, we have identified research directions, from framework-aware analysis and adaptive testing to policy-aware compliance verification and ecosystem-scale comparison, that can guide future efforts.  

By clarifying where current methods encounter limitations and pointing toward ways of combining targeted adaptation with unifying principles, our work aims to provide both researchers and practitioners with a reference point for navigating this evolving field. We hope that this perspective can support the development of more resilient and forward-looking approaches to mobile app analysis, and we look forward to future contributions that build upon and extend these directions.

\section*{Acknowledgment}

This work was supported in part by the National Natural Science Foundation of China (grants No.62572209, 62502168) and the Hubei Provincial Key Research and Development Program (grant No. 2025BAB057).

\bibliographystyle{ACM-Reference-Format}
\bibliography{acmart.bib}

\end{document}